\documentclass[letter]{aa} 
\usepackage{graphicx,amsmath}
\usepackage{txfonts}
\newcommand{\op}{\rm \Omega_p}
\newcommand{\kms}{km s$^{-1}$}
\newcommand{\kmskpc}{km s$^{-1}$ kpc$^{-1}$}
\newcommand{\hi}{\mbox{H{\sc i}}}
\newcommand{\ha}{H$\alpha$}

\newcommand{\fantomm}{\texttt{\textsc{FaNTOmM}}}
\newcommand{\rat}{{\cal R}}
\newcommand{\rbar}{\rm a_B}
\newcommand{\pa}{\rm P.A.}
\newcommand{\as}[2]{#1\arcsec\,\hspace{-1.7mm}.\hspace{.1mm}#2}

\begin{document}
  \title{A slow bar in a dark matter dominated galaxy}


  \author{Laurent Chemin\inst{1} \and Olivier Hernandez\inst{2}
         }

  \institute{G\'EPI-Observatoire de Paris, section Meudon, CNRS \& Universit\'e Paris 7 UMR 8111, 
             5 pl. Janssen, 92195 Meudon, France \\ \email{laurent.chemin@obspm.fr}
        \and
            Universit\'e de Montr\'eal, D\'epartement de Physique, succ. centre-ville, Montr\'eal (QC), Canada H3C3J7\\
            \email{olivier@astro.umontreal.ca}
            }

  \date{Received ; accepted }


 \abstract
{} 
{ We report on an estimate of the bar pattern speed $\rm \Omega_p$ for
the low surface brightness spiral galaxy UGC 628.}
{We applied the Tremaine-Weinberg method to high resolution \ha\ velocity
and integrated emission maps of this dark matter dominated galaxy. 
Observations were made at the CFHT using
the optical Fabry-Perot interferometer, \fantomm.}
{The Tremaine-Weinberg method estimates a bar pattern speed of $(11.3 \pm
2.0)$ \kmskpc\ for UGC 628, which is among the lowest values found for a spiral galaxy.
The corotation radius $\rm R_c$ of the bar and the gaseous disc is $\rm
R_c = 9.8^{+2.9}_{-2.0}$ kpc, implying a ratio $\rm \rat = R_c/\rbar$
of $2.0^{+0.5}_{-0.3}$, where $\rm a_B$ is the bar radius. The ratio is
well beyond the usual range of values, $1.0 \le \rat \le 1.4$, found
for fast bars of high surface brightness barred galaxies.  It implies that the
bar in UGC 628 is slow.}
{As shown through the use of numerical simulations, fast bars survive
when the inner mass distribution of galaxies is dominated by the baryons over the dark matter. 
 Our result suggests that the presence of slow bars in galaxies is likely 
 related to the dominance of dark matter over the mass distribution.}


\maketitle

\section{Introduction}
\label{sec:intro}
In the last decade, estimates of the pattern speed of stellar bars in spiral galaxies have 
shown that they have high amplitudes (Gerssen, Kuijken \& Merrifield 1999; Aguerri, Debattista \& Corsini
2003; Corsini et al. 2007). As a consequence, the ratio of the corotation radius, 
$\rm R_c$, to the semi-major axis of the bar, $\rm a_B$, 
is  usually observed in the range $[1;1.4]$, within the uncertainties. 
A bar whose ratio falls within this range is called a fast bar
 because its pattern speed cannot be larger, according to theoretical 
 models. A bar that would 
 rotate faster would have a smaller corotation radius
 at fixed bar semi-length ($\rat < 1$) and should not exist. 
  Indeed, analytic solutions of the equations of motion 
 and numerical simulations have shown that 
 periodic orbits of stars and gas are aligned perpendicular to the bar 
 in regions beyond corotation and tend to destroy it 
 (Contopoulos \& Papayannopoulos 1980; Athanassoula 1992). 
 The numerical models of Athanassoula (1992) also show simulated bars with $1.0 \le \rat \le 1.4$,  
which result is in excellent agreement with the observations. 
Conversely, stellar bars having $\rat > 1.4$ are called slow bars and can exist, 
though they have not been observed yet.

\label{sec:obs}
  \begin{figure*}[!ht]
  \centering
  \includegraphics[width=\textwidth]{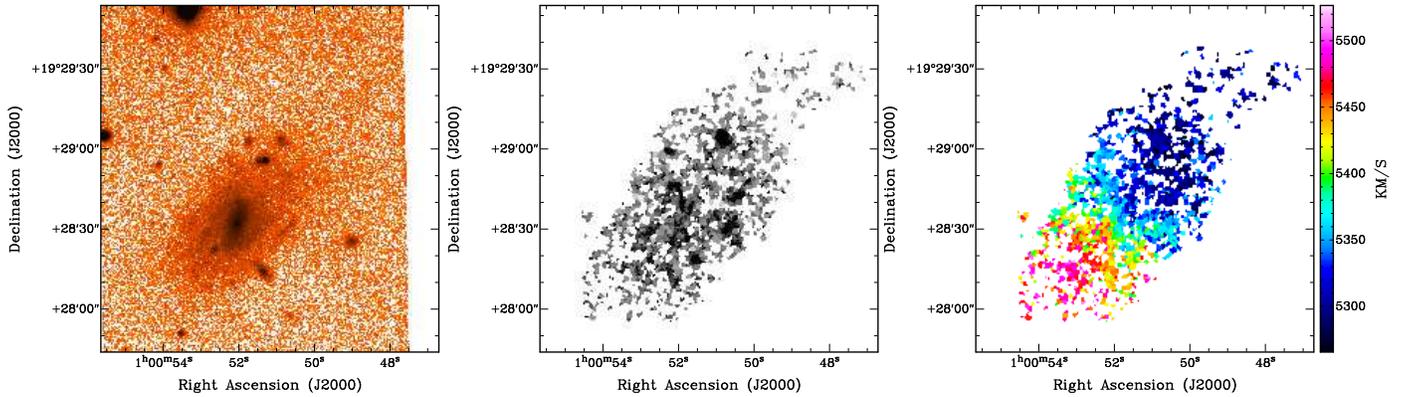}
  \caption{UGC 628. $R-$band image (left panel, de Jong 1996), \ha\ integrated emission and velocity maps (middle and right panels, 
  Chemin et al. in preparation). Logarithmic stretches are used in left and middle panels.}  \label{fig:imu628}%
   \end{figure*}
   
 Classifying a bar as  fast or slow thus relies on the measurements of its
pattern speed, its extent and on the details of the disc rotation curve.
  Most of pattern speed determinations were made through the application of
the Tremaine-Weinberg method (TW; Tremaine \& Weinberg 1984).  Starting
from the continuity equation, Tremaine \& Weinberg showed that a pattern
speed can be defined as $\op \rm \sin(i)= \rm \int^{\infty}_{-\infty}
\Sigma v dx/\int^{\infty}_{-\infty} \Sigma x dx $ where $i$ the disc
inclination, $\Sigma$ is the surface density of the tracer, v is the
line-of-sight velocity and x the position along slices parallel to the
disc major axis.  Since the continuity equation underlies this estimate,
applying the TW method to the old stellar population is most accurate as
it is expected to satisfy the continuity equation better than the gas.
For this reason,  authors have generally used long-slit spectroscopy of
stellar absorption lines of  early type barred discs for these estimates
(Merrifield \& Kuijken 1995; Aguerri et al. 2003; Gerssen, Kuijken \& Merrifield 1999, 2003; Corsini, Debattista \& Aguerri 2003; 
Corsini et al. 2007). Nonetheless, the TW method has been applied with some success to
the kinematics of the neutral, molecular, and warm ionized gas 
(Bureau et al. 1999; Rand \& Wallin 2004; Hernandez et al. 2005; Emsellem et
al. 2006; Fathi et al. 2007; Beckman et al. 2008; Meidt et al. 2008), though 
the conditions under which it can be applied to the gas component are not yet fully understood.  

 In this letter, we estimate the bar pattern speed of UGC 628. 
This galaxy was selected because its morpological 
type and dark matter properties strongly differ from those of galaxies usually observed for 
this measurement. Indeed, UGC 628  is a Magellanic barred, low surface
brightness galaxy (LSB, hereafter). Like most galaxies of this kind,
its mass is entirely dominated by dark matter at all galactocentric radii
(de Blok \& Bosma 2002). 
Our objective is to investigate how the  
 dynamical properties of a bar in a dark matter dominated galaxy compare 
 with those of bars in more typical high surface brightness
galaxies.  

 For this measurement, we used optical 3D spectroscopic observations of
the ionized gas in UGC 628.  It is a difficult task in the emission line
gas and virtually impossible in the stellar absorption lines because
the emission is very faint. This surely explains why pattern speeds have
predominately been measured in high surface brightness
galaxies. 

The kinematical and photometric observations  are described
in Section~\ref{sec:obs} and their analysis is done in
Section~\ref{sec:analysis}.  A discussion and brief concluding remarks
are presented in Section~\ref{sec:discussion}.

\section{Object Characteristics and Observations}

 UGC 628 ($\rm D=71.2$ Mpc, $i=56\degr$) was observed with \textsc{FaNTOmM},
an optical Fabry-Perot interferometer (Hernandez et al. 2003) mounted on 
the Canada-France-Hawaii Telescope (CFHT). These
observations are part of a large survey of the kinematics of low surface
brightness galaxies (see Chemin et al. in preparation for more details).
The angular sampling of the \ha\ datacube is $\as{0}{49}$ per (square)
pixel, corresponding to $\sim 170$ pc at the distance of the galaxy. The
resolving power of the etalon is 12450 at the redshifted \ha\ line.
The spectral range of 262.7 \kms\ was scanned through 24 channels. The
exposure time was 8.1 minutes per channel.  The data reduction procedure
is fully described in previous papers (e.g., Chemin et al. 2006).
Figure~\ref{fig:imu628} shows a $R-$band image of UGC 628 (de Jong
1996), the distribution and the velocity field of the \ha\ emission line.
The broad band image shows a seemingly undisturbed spiral galaxy with an
inner stellar bar aligned approximately North-South. The \ha\ emission
displays a faint disc of diffuse gas. It is brighter in the bar region
as well as along the two spiral arms which are more or less parallel
to the major axis of the disc.  

 \begin{figure}[!b]
  \centering
  \includegraphics[width=\columnwidth]{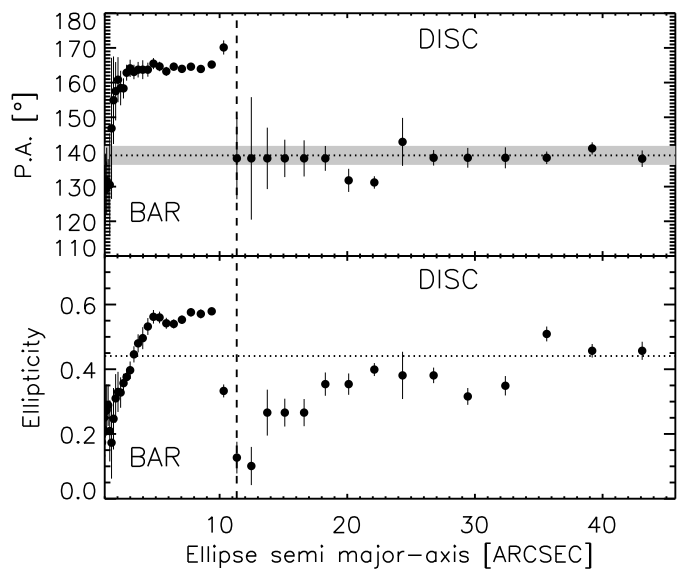}
  \caption{Results of isophotal ellipse fitting of the $R-$band image of UGC 628. The position angle and ellipticity 
  of the isophotes are shown on top and bottom panels respectively as a 
  function on the projected radius. The dashed vertical line indicates the projected bar semi-major axis. The shaded area represents 
  the kinematic major axis of the disc at $\pa = (139 \pm 3) \degr$. The horizontal line in bottom panel is the ellipticity 
  of the disc for an inclination of $56\degr$.}
  \label{fig:ellipse}%
   \end{figure}

\section{Analysis}
\label{sec:analysis}

\subsection{The bar semi-major axis}
\label{sec:ellipse}
 The intrinsic radius of the bar, $\rbar$,  can be estimated following several ways
(Athanassoula \& Misiriotis 2002 and references therein).  We use the
two methods presented in Wozniak et al. (1995) and Aguerri et al. (2000)
to derive $\rbar$ for UGC 628. 

 In the first method, isophotes are fitted to images of the galaxy. We
used the $BVRI$ images from de Jong (1996) and derived profiles of
ellipticity and position angle, $\pa$, of the isophotal major axis using
the \textit{ellipse} task in \textsc{iraf}. The $\pa$ is defined as the
angle between the semi-major axis of the receding half of the velocity
field and North (angle calculated North through East).  Foreground and
clumpy emission regions are masked before fitting.  All photometric
bands give similar results though the $B-$ and $I-$bands show more
scatter which could be due to the low signal-to-noise ratio of the images.
Results given hereafter have been obtained by averaging the values from
all photometric bands.  Only the results for the $R-$band image are shown
for reasons of clarity.  Wozniak et al. (1995) showed that   $\rbar$
can be estimated by determining the location of a discontinuity in the
distribution of $\pa$ and ellipticity with radial distance.  The change
is such the regions in the bar show constant $\pa$ which then abruptly
changes to the disc $\pa$. Similarly, there is an abrupt change in the
ellipticity from a fairly flat distribution to something significantly
more round.   Figure~\ref{fig:ellipse} shows the $R-$band radial profiles
of $\pa$ and ellipticity of UGC 628 in the central regions of the galaxy.
UGC 628 has a very small bulge (Fig.~\ref{fig:imu628}) that should not
contaminate the profiles significantly.  The ellipticity peaks around
a projected radius $r= \as{9}{4}$ before the end of the bar. The bar
position angle is constant on the sky plane ($\sim 163\degr$) and well
separated from the disc orientation ($\pa =$ 138\degr).  The projected
bar radius is $\as{11}{3}$, which corresponds to $\as{13}{2}$ after
deprojection. 

   \begin{figure}[!bh]
  \centering
  \includegraphics[width=\columnwidth]{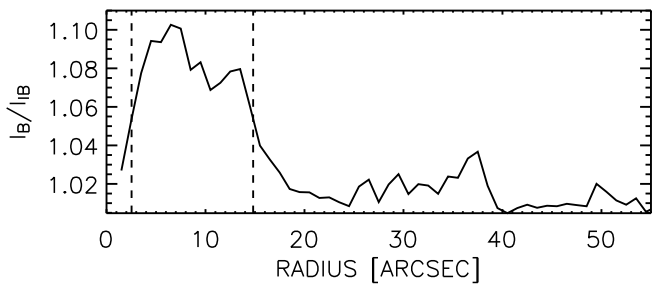}
  \caption{Fourier analysis of a $R-$band image of the LSB galaxy UGC 628. 
   The radial profile of the bar intensity $\rm I_B$ to the interbar intensity $\rm I_{IB}$ is shown. 
   The vertical dashed lines indicate the bar region.  
   The outer line at $R=14.8\arcsec$ corresponds to the deprojected bar semi-major axis.}  \label{fig:fourier}%
   \end{figure}

A second method uses a Fourier decomposition of an image. Aguerri et
al. (2000) define the bar as having the radial range where $\rm I_B/I_{IB}
> 0.5 [ max(I_B/I_{IB}) - min(I_B/I_{IB}) ] + min(I_B/I_{IB})$. In
this expression $\rm I_B = I_0 + I_2 + I_4 + I_6$ is the bar intensity,
with $\rm I_m$ being the even Fourier coefficients of order m, and $\rm
I_{IB} = I_0 - I_2 + I_4 - I_6$ is the interbar intensity.  We fitted
the Fourier coefficients with a least-squares routine to a deprojected
version of the $R-$band image of UGC 628. The results of the Fourier
analysis is shown in Fig.~\ref{fig:fourier}. Following the prescription
of Aguerri et al. (2000), the bar region extends between $\as{2}{5}
\le R \le \as{14}{8}$. The bar radius is the upper limit of this range.

The two bar radius estimates are in very good agreement. The final bar
semi-major axis is derived as the average of these two values, which
gives $\rbar = (14.0 \pm 0.8)\arcsec$ or ($4.8 \pm 0.3$) kpc.

\begin{figure}
\centering
\includegraphics[width=\columnwidth]{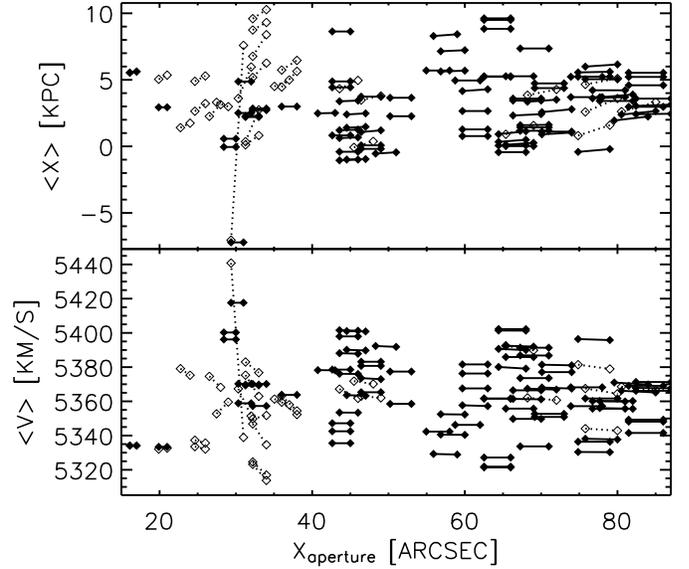}
\caption{Tremaine-Weinberg integrals as a function of the aperture length (or extent of the domain of integration). 
  Filled (open, respectively) symbols and solid (dotted) lines are for the integrals 
  that have (have not) converged to stable values.} 
   \label{fig:convergence}%
   \end{figure}

 \begin{figure}
  \centering
  \includegraphics[width=\columnwidth]{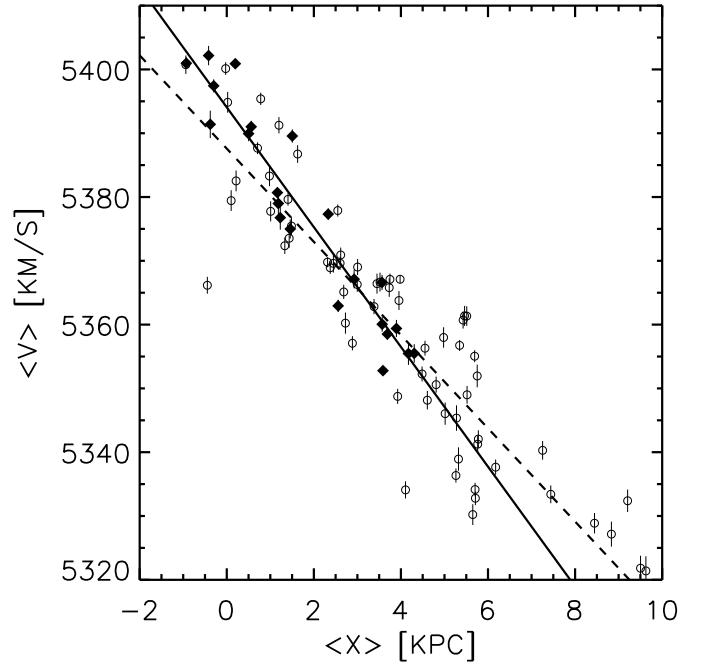}
  \caption{Bar pattern speed measurement of the LSB galaxy UGC 628. Filled (open) symbols are for data inside (outside, respectively) 
  the bar region. The solid line is the best fitting result to the filled symbols, with slope $\op \sin(i)$. The dashed line is the best 
  fitting result to the open symbols, with a slope corresponding to the pattern speed of the spiral arms of UGC 628.}
   \label{fig:twfit}%
   \end{figure}

\subsection{The bar pattern speed}

We use the \ha\ velocity field of UGC 628 (Fig.~\ref{fig:imu628}
and \S\ref{sec:obs}) with which intensity weighted mean positions and
velocities are measured along many slices aligned parallel to the kinematical major
axis of the galaxy. This is the TW method.

 Though reasonable values of pattern speeds are derived using \ha\
observations (e.g., Beckman et al. 2008), the conditions under which the
method is applicable to the warm ionized gas is not entirely clear.  
 However, Hernandez et al. (2005), through the use N-body/SPH
simulations found that the method was robust for the ionized gas but
with proviso that prominent regions where strong gas shocks occur should
be avoided.  Such regions introduce mean local velocity and position
discontinuities that can bias slope measurements used in TW method.
Fortunately, such discontinuites are easily identifiable in the TW
diagrams.  

The kinematical position angle of the disc ($\pa = (139 \pm 3)\degr$)
defined the orientation of the slices.  It is derived from 
  a tilted-ring model of the \ha\ velocity field (Chemin et al., in prep.). There is
very good agreement between the photometric and kinematical major axes
(\S\ref{sec:ellipse}). The pattern speed of the bar is taken to be
constant as a function of radius so that a single value can be derived
using the TW equation of \S\ref{sec:intro}.  We refer to Merrifield,
Rand \& Meidt (2006) and Meidt et al. (2008) for discussions of the
radial dependence of pattern speed measurements in galaxies.

 In practice, the integrals of $\rm \Sigma v$ and $\rm \Sigma x$
are not done within $]-\infty,+\infty[$ because of the finite extent of
the galaxy due to the surface brightness detection limit. The integrals
are calculated along slices of maximal aperture $\rm X_{max}$, with
the value of $\rm X_{max}$ differing from one slice to another one.
Results for two apertures values $\rm X_{aperture} = X_{max} - \epsilon$
and $\rm X_{aperture} = X_{max}$  are shown in
Figure~\ref{fig:convergence}, where $\epsilon$ is a small fraction of
$\rm X_{max}$. Most of the integrals have converged
(or nearly so) to stable values $<$X$>$ and $<$V$>$
at these two apertures (filled symbols and solid lines). 
The TW method thus seems to be applicable to the ionized gas data of UGC 628.
Integrals that
have not converged within $\rm X_{max}$ are displayed with open symbols
and dotted lines.  They correspond to the edges of the gas distribution
and velocity field, i.e. the most distant regions to the major axis. 
This is not surprising because the slices in these regions contain very few points. 

Figure~\ref{fig:twfit} shows the result of applying the Tremaine-Weinberg
method to UGC 628.   Only integrals which have converged
within $\rm X_{aperture} = X_{max}$ are drawn and kept in the slope fitting of the relationship.
The uncertainties in the velocities are typically of the order of 3-4
\kms. They depend mainly on the accuracy of the wavelength calibration of
the raw datacube.  The accuracy on the velocity centroid of an emission
line in each pixel is indeed a fraction of a velocity channel, as derived from
observations of a reference Neon lamp.   

 Due to the lopsided galaxy morphology, the average positions are preferentially observed 
at positive offests. A similar effect has already been observed in other \ha\ or CO observations (e.g., 
 Rand \& Wallin 2004; Hernandez et al. 2005;  Emsellem et al. 2006). 
As a check for consistency with the stellar tracer, $<$X$>$ values have been 
derived from the $R-$band image. They have been observed to be offset towards positive values as well. 

Only points which location falls
within the bar region $\rm R \le 14\arcsec$ in the 2D \ha\ image are used
to measure the bar pattern speed (filled symbols).
 The best fit gives
$\op = 11.3$ \kmskpc.  This value does not change when a larger region is
used, $\rm R \le \as{14}{8}$ and only increases by $2\%$ when a smaller
region is used, $\rm R \le \as{13}{2}$.  All other points lying beyond
the bar extent (open symbols) are considered to trace another pattern speed (e.g., that 
of  the spiral arms in the disc). An amplitude of 
8.8 \kmskpc\ is estimated for them, which value is lower than the bar pattern speed.

The total uncertainty on $\op$ is obtained by adding in quadrature the
uncertainties in $\op$ due to the uncertainty in the position angle
(3\degr), the inclination (2\degr), the extent of the region used
for fitting $\op$ (which is the error on the region of influence of
the bar, i.e., fixed at the bar radius uncertainty of $\as{0}{8}$)
and to the formal $1\sigma$ uncertainty in the fit. All of the three
later uncertainties are of-order $0.1-0.2$ \kmskpc, which is negligible
compared to uncertainty due to the uncertainty in the position angle.
3D spectroscopy allows us to derive $\op$ at different position angles
(between 136\degr\ and 142\degr) and we found a significant change of
$\sim$18\% in $\op$ ($\sim$2.0 \kmskpc). $\op$ increases (decreases)
towards smaller (larger) position angles than $\pa = 139\degr$. This is
similar to the effect seen by other authors (Debattista 2003; Debattista
\& Williams 2004).  Because of this dominant source of uncertainty,
we adopted $\op = (11.3 \pm 2.0)$ \kmskpc\ as the likely best value.

\section{Discussion and conclusion}
\label{sec:discussion}

\begin{figure}
  \centering
  \includegraphics[width=\columnwidth]{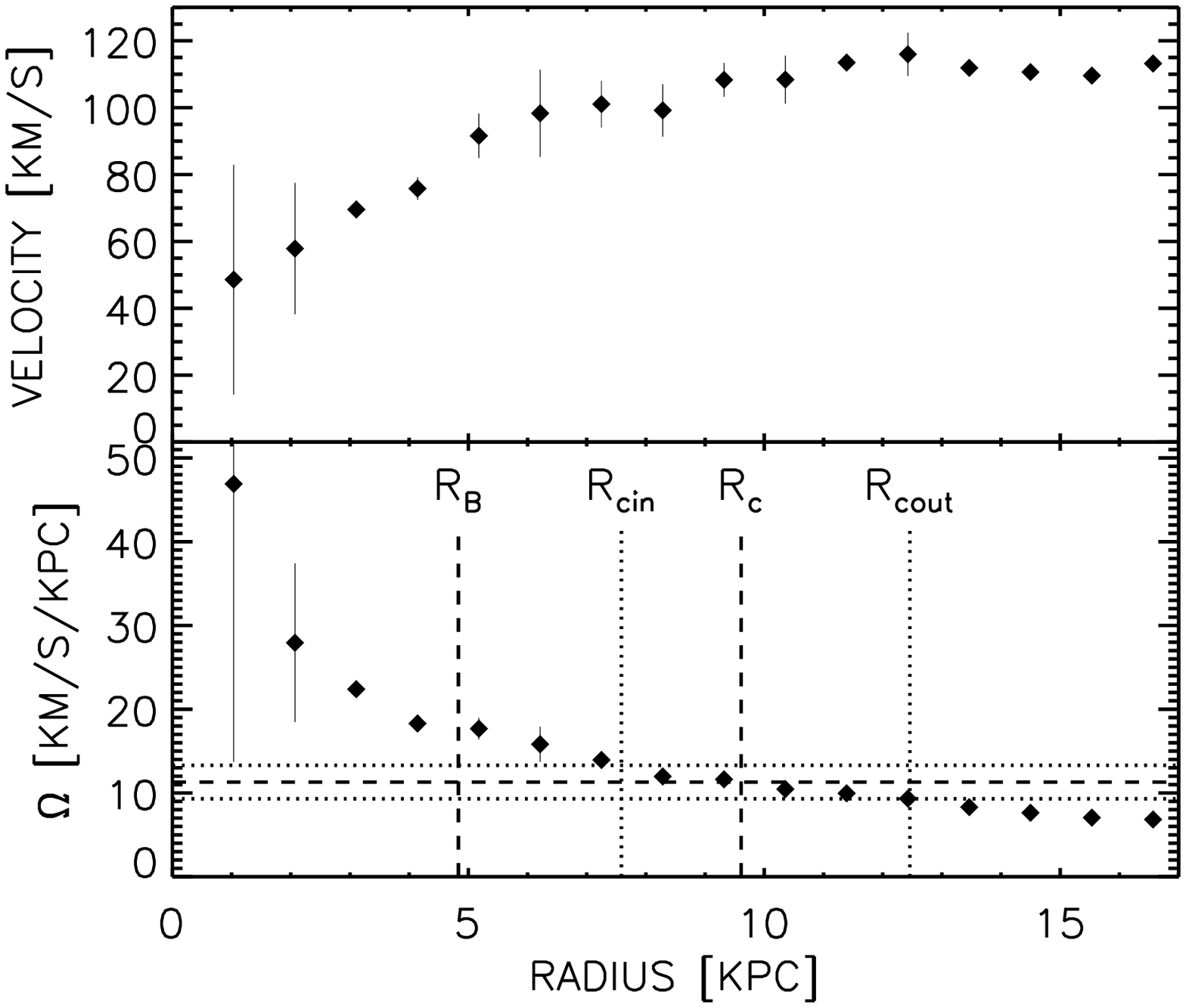}
  \caption{Rotation curve (top panel) and angular velocity curve (bottom panel) of the LSB galaxy 
  UGC 628. Dashed and dotted horizontal lines are the measured bar pattern speed of 
  $\op = (11.3 \pm 2.0)$ \kmskpc. Dashed and dotted vertical 
  lines respectively indicate the bar semi-major axis $\rm a_{B}$, the corresponding corotation radius $\rm R_c$ and its lower and upper limits ($\rm R_{cin}, R_{cout}$) 
  allowed by the uncertainties on $\op$. }
   \label{fig:rc}%
   \end{figure}

 Figure~\ref{fig:rc} (top panel) shows the \ha\ rotation curve of
UGC 628 (Chemin et al. in prep.), derived using a tilted-ring model
of the velocity field. The angular velocity profile $\rm \Omega(R)=
\frac{V(R)}{R}$ (bottom panel) was used to determine the corotation
radius of the bar with the disc ($\Omega = \op$). The corotation radius
is $\rm R_c = 9.8^{+2.9}_{-2.0}$ kpc implying a ratio $\rm \rat =
R_c/\rbar = 2.0^{+0.5}_{-0.3}$.  
Given that observed  bars having $1.0 \le \rat \le 1.4$ are defined as being fast 
 (e.g., Gerssen et al. 1999) and larger values of $\rat$ correspond to slow bars, 
  UGC 628 appears to host a slowly rotating bar. 

 The slowest bar found before UGC 628 was for the blue compact dwarf galaxy
NGC 2915 having $\rat > 1.7$  (Bureau et al. 1999).  What is surprising here is
that the bar-like structure of NGC 2915 is only seen in its neutral hydrogen distribution, 
not in the distribution of its stellar light. Purely gaseous bars are very rare in galaxies because 
bars are generally devoid of gas. Apart from that noteable exception, fast
bars are always seen in galaxies (Gerssen et al. 1999, 2003; Aguerri et al. 2003; Debattista \&
Williams 2004; Corsini et al. 2007). To our knowledge, UGC 628 is thus the first case of 
an obvious slow stellar bar in a galaxy.

A striking point is that all fast bars are found in normal high surface
brightness early-type discs. The stellar mass dominates the dark matter
mass in their central regions, their disc is thought to be maximum.
It is known that low surface brightness  discs are dark
matter dominated at all galactocentric distances, even with realistic
mass-to-light ratio for their stellar disc (de Blok \& Bosma 2002).
From that, it seems that minimum disc galaxies tend to host slow bars,
in opposition to what is observed in maximum disc galaxies.  Moreover,
the fact that UGC 628  is at the opposite side of the Hubble diagram
to the bright galaxies is perhaps the reason why it is observed a slow
pattern speed for it.  With two cases of slow stellar and gaseous bars in
UGC 628 and NGC 2915 (respectively), it is thus tempting to argue that
the amplitude of bar pattern speeds decreases as a function of the Hubble type. 
This is in agreement with results from N-body simulations (Combes \& Elmegreen 1993).

Other numerical simulations explain how fast bars can be formed and maintained in a 
high surface brightness galaxy, preferentially within a maximum disc (Debattista \& Sellwood 2000; Valenzuela \& Klypin 2003). These simulations 
are in agreement with many observations of pattern speeds. 
 Debattista \& Sellwood (2000) argue that barred galaxies with moderately dense halos (like LSBs) should host a slow bar. 
 Our result is therefore in full agreement with their expectation.
 Other numerical works investigate the bar evolution in LSB discs embedded in cosmological NFW halos 
(Mayer \& Wadsley 2004). However they do not address the question of the bar pattern speed and the bar length compared with the corotation 
radius. 

 Additional  works are needed to confirm this new result. 
First, additional observations of barred low surface brightness discs 
would help to conclude whether UGC 628 is an isolated case or whether slow bars are common in 
this type of galaxies. They would also help to investigate whether pattern speeds really 
decrease as a function of the morphological type. 
High-resolution \hi\ or optical 
interferometry like the current Fabry-Perot measurement seems to be 
the key for the Tremaine-Weinberg method because it is very difficult to observe stellar absorption lines  
of LSBs and the molecular gas is hardly detectable in them (O'Neil \& Schinnerer 2004). 
A careful analysis of the datacubes of barred LSBs from our sample will be done. 

 Finally, additional numerical simulations would also be very helpful to 
investigate the properties of bars in  low surface density, dark matter dominated galaxies. 
Low  mass surface density discs are responsible of the cusp-core controversy (de Blok \& Bosma 2002; Hayashi, Navarro \& Springel 2007). 
Kinematical observations of LSBs indeed show that their dark halo seem to have a constant density in their core while steeper density profiles 
are expected in Cold Dark Matter numerical simulations (e.g. Navarro, Frenk \& White 1997). 
How would the pattern speed of a bar embedded within a cusp compare with that of a bar embedded within a core halo? 
Such simulations could perhaps help in constraining the shape of the inner density profile  and alleviate 
the cusp-core controversy.

\begin{acknowledgements}
We are very grateful to  P. Amram, C. Balkowski and C. Carignan for their
long term support in this project and to M. D. Lehnert for fruitful discussions and a careful reading of the 
manuscript.  We acknowledge an anonymous referee for very constructive remarks. 
This work is based on observations
collected at the Canada-France-Hawaii Telescope, which is operated by the
National Research Council of Canada, the Centre National de la Recherche
Scientifique de France and the University of Hawaii.  OH thanks the Fond
Qu\'eb\'ecois de la Recherche sur la Nature et les Technologies and the
National Research Council of Canada for their support.
\end{acknowledgements}

\end{document}